\documentclass[final,1p,times]{elsarticle}
\usepackage{graphicx} 
\usepackage{amssymb} 
\usepackage{amsthm} 
\usepackage{lineno} 

\journal{Nuclear Physics A} 
\begin{document} 

\begin{frontmatter}

\title{A Reanalysis of Single Photon Data at CERN SPS}

\author{Charles Gale$^{ a,}${\footnote {Speaker, Quark Matter 2009 (QM09), 
March 30-April 4, 2009, Knoxville, TN, USA.}}, Rupa Chatterjee$ ^b$, 
Dinesh K. Srivastava$ ^b$, and Sangyong Jeon$ ^a$}

\address{$^a$Department of Physics, McGill University, Montreal, Canada H3A 2T8
\\ $^b$ Variable Energy Cyclotron Centre, 1/AF Bidhan Nagar, Kolkata 700 064, India}

\begin{abstract} 
We reanalyze the WA98 single photon data~\cite{wa98}
at CERN SPS by incorporating several recent developments 
in the study of prompt and thermal photon production 
from relativistic heavy ion collisions~\cite{csj}. 
Isospin and shadowing corrected NLO pQCD, along with 
an optimized scale for factorization, fragmentation 
and renormalization are considered for prompt photon 
production. Photons from thermal medium are estimated 
by considering a boost invariant azimuthally anisotropic 
hydrodynamic expansion of the plasma along with a well 
tested equation of state and initial conditions. 
A quantitative explanation of the data is obtained by 
combining $\kappa \, \times$ prompt with thermal photons, 
where $\kappa$ is an overall scaling factor. We show that, 
elliptic flow of thermal photons can play a crucial role to 
distinguish between the `with' and `without' phase 
transition scenarios at SPS energy.
\end{abstract} 

\end{frontmatter} 

\section{Introduction}

The first observation of single photons by the WA98 Collaboration 
at CERN SPS is considered as a well anticipated turning point 
in the study of relativistic heavy ion collisions using 
electromagnetic probes~\cite{wa98}. Earlier observations 
like the one by the WA80~\cite{wa80} Collaboration, provided 
only a useful upper limit of this study (for recent 
developments in the field of direct photon production from 
relativistic heavy ion collisions, see Ref.~\cite{stankus, dks}). 
The study of electromagnetic radiations, and in particular 
photons, as a probe of heavy ion collisions is advantageous 
compared to the study of hadrons, for two main reasons. 
First of all, photons are emitted from each and every stage 
of the expanding system, whereas hadrons are emitted only 
from the surface of freeze-out after suffering strong 
interactions. Secondly, photons do not suffer final state 
interaction (for being electromagnetic in nature, their 
mean free path is larger than the system size) and carry 
undistorted information from the production point to the 
detector. The major problem in the study of single photons 
from heavy ion collisions arises from the very small signal 
to background ratio. However, recent developments in the 
background subtraction methods have reduced the size of error 
bars in the direct to decay ratio for photons considerably.
We reanalyze the single photon WA98 data by incorporating 
several new improvements in our understanding of prompt 
photon production from heavy ion collisions and considering 
the latest developments in the field of thermal photon 
production along with a well defined equation of state and 
suitable initial conditions~\cite{csj}.

\section{Reanalysis of single photon data at CERN SPS}
\subsection{Prompt photons}
The study of prompt photon production in $p+p$ collisions has 
reached a higher level of sophistication and all the available 
data have now been successfully analyzed with NLO pQCD 
treatment~\cite{nlo} without the  inclusion of intrinsic $k_T$ 
for protons. In particular, the suppression of single 
photons at large $p_T$ for Au+Au collisions with respect 
to the single photons resulting from $p+p$ collisions at the 
same nucleon-nucleon center of mass energy may be largely due 
to the difference in valence quark structure of protons and 
neutrons \cite{iso}. 
\begin{figure}[t]
\begin{minipage}{15pc}
\includegraphics[width=10.5pc,angle=-90,clip=true]{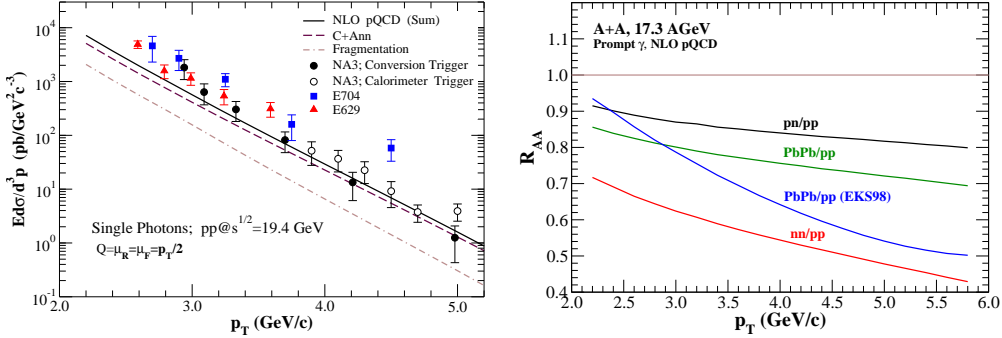}
\end{minipage}\hspace{1pc}%
\begin{minipage}{15pc}
\includegraphics[width=15pc,clip=true]{iso.eps}
\end{minipage} 
\caption{\label{pp}[Left panel] Prompt photons from p+p 
collisions at $\sqrt{s}$ = 19.4 GeV. Experimental results 
for E704~\cite{e704} for p+p collisions and those estimated 
from $p+^{12}C$ collisions by E629~\cite{e629} and 
NA3~\cite{na3} experiments are given for a comparison. 
[Right panel] Effect of isospin and parton shadowing on 
the production of prompt photons at $\sqrt{s_{NN}}$ = 17.3 
GeV.}
\end{figure}

We calculate the prompt photon production from $p+p$ collisions 
using NLO pQCD treatment along with an optimized scale for 
factorization, fragmentation and renormalization (all equal 
to $p_T/2$) at $\sqrt{s} \ = \ 19.4$ GeV (without introducing 
intrinsic $k_T$) and compare our results with various 
experimental data available at that energy (with proper 
mass number normalization for $p+ ^{12}C $ collisions). 
This comparison is done as no experimental data are available 
for $p+p$ or $n+n$ collisions at the WA98 center of mass 
energy ($\sqrt{s_{NN}} \ = \ 17.3$ GeV) and the available 
data at the closest center of mass energy is at $\sqrt{s} 
\ = \ 19.4$ GeV [see left panel of  Fig.~\ref{pp}]. We see 
that the photons originating from fragmentation are about 30\% 
of those produced from Compton+annihilation processes. Our 
result using NLO pQCD matches well with the NA3~\cite{na3} data, 
while it underestimates the E704~\cite{e704} and E629~\cite{e629} 
data, same as reported by earlier studies.

For prompt photon production from 158A GeV  Pb+Pb collisions, 
isospin and shadowing \cite{eks} corrected NLO pQCD treatment is used 
with the same scaling factor of $p_T/2$ for factorization, 
fragmentation and renormalization. The effect of isospin and 
shadowing on photon production from heavy ion collisions are 
investigated  by calculating nuclear modification factor ($R_{AA}$) 
as function of $p_T$ and $x_T \ ( \ = 2p_T/\sqrt{s} \ )$ for 
different beam energies. Results for p+p normalized $R_{AA}$ as 
a function of $p_T$  for p+n, n+n, and Pb+Pb collisions are shown 
in right panel of Fig.~\ref{pp}. We see that the photon production 
from Pb+Pb collisions is suppressed significantly in the 
intermediate and high $p_T$ range, compared to the production 
from p+p collisions.
\begin{figure}[t]
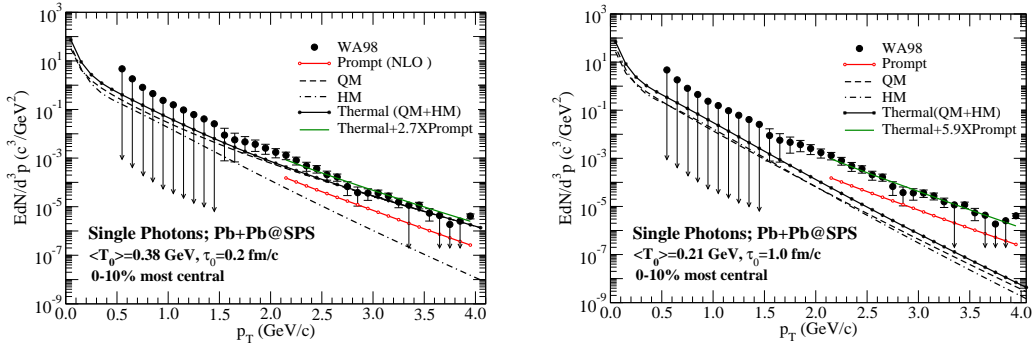

\begin{minipage}{15pc}
\includegraphics[width=15pc,clip=true]{pb_0.2.eps}
\end{minipage}\hspace{1pc}
\begin{minipage}{15pc}
\includegraphics[width=15pc,clip=true]{pb_1.0.eps}
\end{minipage}\hspace{1.2pc}
\caption{\label{sps}Fit to single photon spectra from
Pb(158A GeV)+Pb collisions measured by the WA98~\cite{wa98} 
experiment for $\tau_0 = 0.2 $ fm/$c$ [left panel] and 
1.0 fm/$c$ [right panel] using scaling factor $\kappa =$ 
2.7 and 5.9 respectively.}
\end{figure}
\subsection{Thermal photons}
For thermal photons, centrality dependent azimuthally anisotropic 
boost invariant ideal hydrodynamics is used along with 
different sets of initial parameters. A well tested equation of 
state is used considering a first order phase transition from the 
plasma state to the hadronic phase at a transition temperature 
$T_c \ ( \sim 164)$ MeV. The initial energy density profile is 
taken as proportional to the number of wounded nucleons and five 
different values of $\tau_0$ are  considered ranging from 0.2 fm/$c$ 
to 1.0 fm/$c$ (in steps of 0.2 fm/$c$) keeping the total entropy 
of the system fixed. The time evolution of average energy density 
$\langle \epsilon \rangle$, average temperature $\langle T \rangle$ 
and average radial flow velocity $\langle v_T/c \rangle$ at different 
$\tau_0$ are compared. 
We find that the values of $\langle \epsilon \rangle$ ($\sim T^4$ ) 
changes significantly at large $\tau$ with changing values of 
$\tau_0$, whereas $\langle T \rangle$ and $\langle v_T/c \rangle$ 
are not affected much [see Ref.~\cite{csj} for detail]. Also, the 
effective temperature, $T_{\rm eff}= T \sqrt{ (1+v_T)/(1-v_T) }$, 
(or the blue shifted temperature) is calculated as function of 
proper time at different $\tau_0$ to see the combined effect of 
cooling and expansion (velocity). Thermal photons at different 
$\tau_0$ are calculated considering standard rates (QGP photons 
from Ref.~\cite{AMY}, and HM photons from Ref.~\cite{TRG}) of 
photon production for 0-10\% most central collisions with freeze-out 
energy density of about 0.075 GeV/fm$^3$.

We find that the prompt photon production is about 17\% of the 
total yield measured by WA98 and the thermal photon result is 
almost similar to prompt photon production at $\tau_0 \ = \ 0.4$ 
fm/$c$. We also note that the thermal photons from hadronic 
phase are not affected significantly with changing $\tau_0$. 
A quantitative description of the WA98 experimental data is 
obtained by using the relation $`{\rm Thermal} + \kappa \times 
{\rm Prompt}'$ where, $\kappa$ is adjusted to reproduce the 
photon production at $p_T = 2.55$ GeV/$c$. For all $\tau_0$, 
a normalization at the same $p_T$ (= 2.55 GeV/$c$) provides 
a good description of the data in the entire $p_T$ range. We 
find that the scaling factors $\kappa =$ 2.7, 4.9, 5.4, 5.7, 
and 5.9 for prompt photons at $\tau_0 =$ 0.2, 0.4, 0.6, 0.8, 
and 1.0 fm/$c$  respectively provide a good quantitative 
agreement with the WA98 data [shown in Fig.~\ref{sps}]. We 
argue that the factor $\kappa$ for prompt photons accounts for   
the Cronin effect, in the case of nucleus-nucleus collisions, 
as well as a pre-equilibrium contribution which must surely be 
included when $\tau_0$ is large.

\subsection{Elliptic flow of photons and hadrons}
\begin{figure}[t]
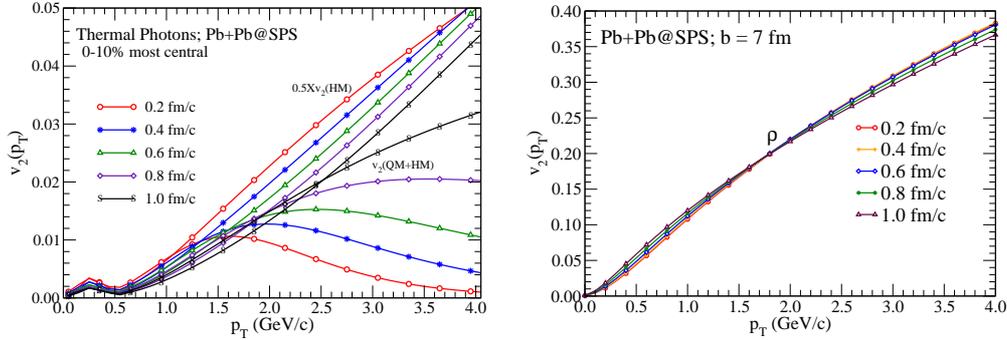

\begin{minipage}{15pc}
\includegraphics[width=15pc,clip=true]{all_v2.eps}
\end{minipage}\hspace{1pc}
\begin{minipage}{15pc}
\includegraphics[width=15pc,clip=true]{v2_rho.eps}
\end{minipage} 
\caption{\label{v2} $v_2$ for thermal photons for different 
$\tau_0$, along with  contributions from hadronic matter. 
[Right panel] $v_2(p_T)$ for primary $\rho$ mesons from 
Pb + Pb collisions having b = 7 fm at SPS energy, for 
different $\tau_0$.}
\end{figure}

In a potentially interesting observation we show that, one 
additional experimental result, i.e, the elliptic flow  
for thermal photons~\cite{cfhs} could actually distinguish 
between the different values of $\tau_0$. The elliptic flow 
results for different $\tau_0$ along with the hadronic matter 
contribution for 158A GeV Pb+Pb collisions at CERN SPS are 
shown in right panel of Fig.~\ref{v2}. We note that several 
earlier studies have explained the WA98 data considering only 
the formation a hot hadronic gas in the collision and without 
the formation of QGP phase~\cite{jane}. The estimation of 
photon flow at SPS can distinguish between the two scenarios 
of `with' and  `without' phase transitions as the nature of 
$v_2$ would be completely different in the two cases. We also 
compute the particle spectra and $v_2(p_T)$ for several hadrons 
and show explicitly that both the spectra and elliptic flow 
results remain unaffected with changing values of $\tau_0$ for 
hadrons. $v_2(p_T)$ for $\rho$ mesons at different $\tau_0$ are 
shown in right panel of Fig.~\ref{v2}.

In conclusion, we present a quantitative explanation of the 
WA98 single photon data at CERN SPS by incorporating several 
recent developments in the field of prompt and thermal photon 
production from heavy ion collisions. Thermal photons at 
different $\tau_0$ along with prompt contribution enhanced 
by a `$\kappa$' factor describes the data quite well in the 
entire $p_T$ range. We also show that thermal photon $v_2$ 
can distinguish between the different $\tau_0$ and phase 
transition scenarios at SPS energies.

\section*{Acknowledgments} C. G. and S. J. acknowledge funding 
by the Natural Sciences and Engineering Research Council of Canada. 
D. K. S. would like to acknowledge the warm hospitality at McGill 
University, where part of the work was done under the McGill India 
Strategic Research Initiative.

\end{document}